\documentclass[10pt,aps,prx,twocolumn,notitlepage,showpacs,superscriptaddress,longbibliography]{revtex4-2}

\usepackage{times}
\usepackage{amssymb,amsmath}
\usepackage{bm}
\usepackage{graphicx}
\usepackage{xcolor}

\usepackage[urlcolor=blue,colorlinks=true,citecolor=blue,linkcolor=blue,pdfstartview={FitH},bookmarks=false]{hyperref}
\begin{document}
	
\title{High-Temperature Superconductivity  from  Finite-Range Attractive Interaction}

\date{\today}
\author{Dmitry Miserev,$^{1\ast}$ Joel Hutchinson,$^1$ Herbert Schoeller,$^2$ Jelena Klinovaja,$^{1}$ and Daniel Loss}
\affiliation{Department of Physics, University of Basel, 
	Klingelbergstrasse 82, CH-4056 Basel, Switzerland\\
	$^2$Institut f\"{u}r Theorie der Statistischen Physik, RWTH Aachen University and JARA -- Fundamentals of Future Information Technology,\\
	52056 Aachen, Germany}

\begin{abstract}
	In this letter we consider $D$-dimensional interacting Fermi liquids, and demonstrate that an attractive interaction with a finite range $R_s$ that is much greater than the Fermi wavelength $\lambda_F$ breaks the conventional BCS theory of superconductivity.
	In contrast to the BCS prediction of a finite superconducting gap for all attractive contact interactions, we show that a finite-range interaction does not induce a superconducting gap. 
	Instead, the pair susceptibility develops a power-law singularity at zero momentum and zero frequency signaling quantum critical behavior without long-range ordering.
	Starting from this, we show that superconductivity can be stabilized by adding a short-range attractive interaction, which is always present in real electronic systems.
	As an example, we consider a layered quasi-two-dimensional material with attractive electron-electron interactions mediated by  optical phonons. We demonstrate a dome shape of the critical temperature $T_c$ versus doping, strongly suppressed isotope effect, and a weak dependence of the optimal doping and maximal $T_c^* \sim 0.1 E_F$ on the interaction range at $R_s \gg \lambda_F$, $E_F$ is the Fermi energy. 
	We believe that these results could be relevant to high-temperature superconductors.
\end{abstract}

\maketitle	

\textit{Introduction.}
BCS and closely related Eliashberg theories provide a clear microscopic mechanism of superconductivity (SC) in metals, predicting a finite SC gap for an arbitrary attractive interaction \cite{bardeen1957,Eliashberg1960,chubukovEliashbergTheoryPhononmediated2020,sharmaSuperconductivityCollectiveExcitations2020}.
In this letter, we show that the BCS approximation breaks down if the interaction range $R_s$ is much greater than the average inter-electron distance characterized by the Fermi wavelength $\lambda_F$.
Surprisingly, such finite-range interactions do not instigate SC order, yet the pair susceptibility develops a power-law singularity at zero momentum and zero frequency, a clear signature of quantum critical behavior~\cite{miserevMicroscopicMechanismPair2024}.
We further show that SC is stabilized by the residual short-range interaction.

As an example, we consider a layered three-dimensional (3D) metal where electrons are restricted to move within weakly coupled two-dimensional (2D) planes.
Attraction between electrons is mediated by an optical phonon and can be described by the effective electron-phonon interaction (EPI) with a finite range $R_s \gg \lambda_F$, where $R_s$ stands for the screening length of the Coulomb interaction.
We find the following results: (i) a dome-shape density dependence of the critical temperature $T_c$, (ii) strongly suppressed isotope effect, (iii) weak dependence of the maximal $T_c^* \sim 0.1 E_F$ and the optimal electron density on $R_s$ as soon as $R_s \gg \lambda_F$, $E_F$ is the Fermi energy.
These features have been observed in a large variety of quantum materials encompassing high-temperature SCs, heavy-fermion materials, magic-angle twisted bilayer graphene and other quantum materials \cite{crawfordOxygenIsotopeEffect1990,franckCopperOxygenIsotope1993,pringleEffectDopingImpurities2000,zhaoUnconventionalIsotopeEffects2001,fournierInsulatorMetalCrossoverOptimal1998,huckerCompetingChargeSpin2014,satoThermodynamicEvidenceNematic2017,xieSpectroscopicSignaturesManybody2019,wongCascadeElectronicTransitions2020,luSuperconductorsOrbitalMagnets2019,yankowitzTuningSuperconductivityTwisted2019,caoUnconventionalSuperconductivityMagicangle2018,ohEvidenceUnconventionalSuperconductivity2021,parkTunableStronglyCoupled2021,aroraSuperconductivityMetallicTwisted2020,wangCorrelatedElectronicPhases2020,shenCorrelatedStatesTwisted2020,rickhausCorrelatedElectronholeState2021,suSuperconductivityTwistedDouble2023,barrierCoulombScreeningSuperconductivity2024}.

\begin{figure}[t]
	\centering
	\includegraphics[width=0.99\columnwidth]{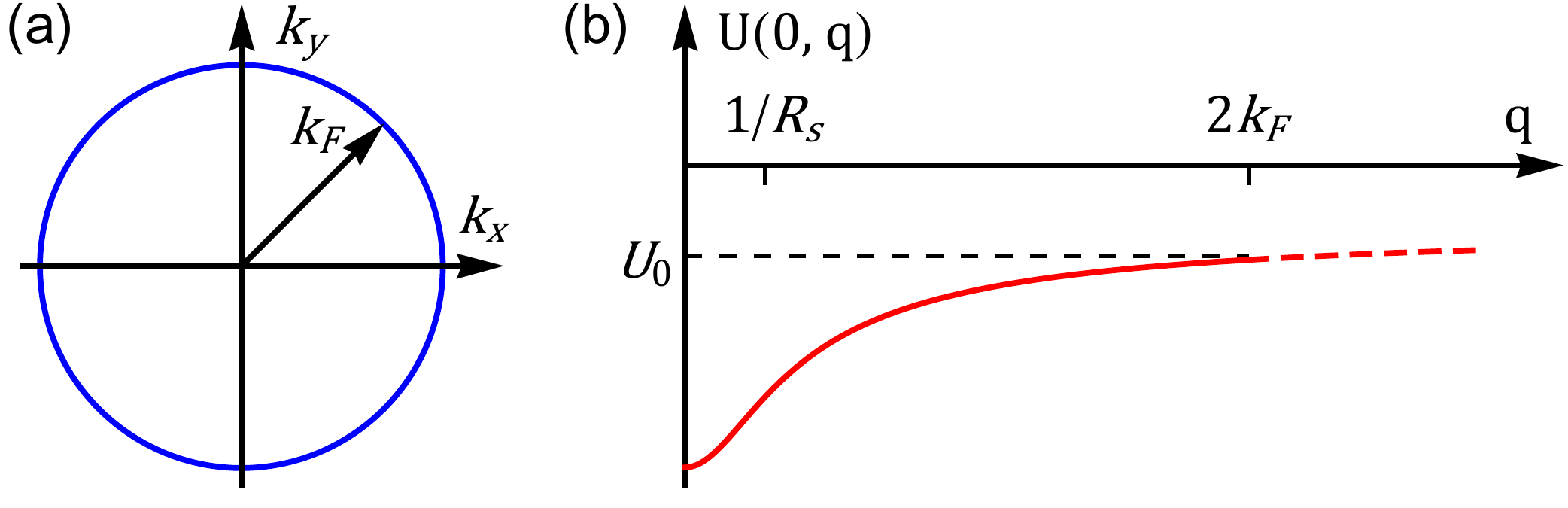}
	\caption{(a) $D$-dimensional isotropic Fermi surface of radius $k_F$, $D > 1$.
	(b) Attractive interaction $U(0, q) = U_0 + U_f(0, q)$ at zero frequency. The constant part corresponds to the short-range matrix element $U_0$ transferring all momenta $q \in [0, 2 k_F]$.
	The finite-range (non-constant) part of the interaction, $U_f(0, q)$, is localized to small momenta $q \lesssim 1/R_s \ll k_F$, where $R_s$ is the screening length (interaction range). The interaction at $q > 2 k_F$ (red dashed curve) scatters electrons off the Fermi surface and therefore is irrelevant.}
	\label{fig:1}
\end{figure}

\textit{Theoretical model.}
First, we present a general theory for a spin-degenerate $D$-dimensional electron gas ($D > 1$) with an isotropic Fermi surface with the Fermi momentum $k_F = 2 \pi/ \lambda_F$, see Fig.~\ref{fig:1}(a).
The attractive interaction $U(\tau, \bm r) = U_f (\tau, \bm r) + U_0 (\tau, \bm r)$ is separated into the finite-range component $U_f(\tau, \bm r)$ acting on large distances $R_s \gtrsim r \gg \lambda_F$, and the short-range component $U_0(\tau, \bm r)$ acting on short distances $r \lesssim \lambda_F$. Here, we assume that $U(\tau, \bm r) = U(\tau, r)$.
A sketch of the zero-frequency Fourier transform $U(0, q)$ is shown in Fig.~\ref{fig:1}(b), where $U_0$ transfers all momenta $q = |\bm q| \in [0, 2 k_F]$, while $U_f(0, q)$ transfers small momenta $q \lesssim 1/R_s \ll k_F$.
Momentum transfers of $q > 2 k_F$ result in scattering off the Fermi surface. Such processes are not resonant and therefore can be omitted.

The mean-field Hamiltonian $H_0$ written in the Nambu representation takes the form, $H_0 = \xi_{\bm k} \eta_z + \Delta \eta_x$, where $\xi_{\bm k} \approx v_F (k - k_F)$ is the electron dispersion linearized near the Fermi surface, $v_F$ is the Fermi velocity and $k=|{\bm k}|$,  $\Delta$ is the SC gap, $\eta_{x,y,z}$ are the Pauli matrices acting on the particle-hole (Nambu) subspace where the field operators are represented by $\Psi_{\bm k}^T = (c_{\bm k, \uparrow}, c_{-\bm k, \downarrow}^\dagger)$, and $c_{\bm k, \sigma}$ ($c_{\bm k, \sigma}^\dagger$) corresponds to the annihilation (creation) operator with momentum $\bm k$ and spin $\sigma \in \{\uparrow, \downarrow\}$.

We study SC using the zero-momentum static pair susceptibility $\chi = \chi(T, \Delta)$ representing the propagator of the Goldstone Cooper pair mode \cite{kulikPairSusceptibilityMode1981},
\begin{eqnarray}
	&& \hspace{-15pt} \chi \! = \!\! \int\limits_0^\beta \!\! \frac{d\tau}{2\Omega}  \sum\limits_{\bm k, \bm p} \!  \mathrm{Tr}\left[ \left\langle \mathcal{T} \! \left\{\! \left(\Psi^\dagger_{\bm k} \eta_y \Psi_{\bm k} \right)\! (\tau) \left(\Psi^\dagger_{\bm p} \eta_y \Psi_{\bm p} \right)(0)  \right\} \right\rangle \right] \! , \label{definitionchi}
\end{eqnarray}
where $\bm k$ and $\bm p$ are $D$-dimensional momenta, $\mathcal{T}$ stands for the time ordering, $\langle \dots \rangle$ corresponds to an average over the statistical ensemble, $\mathrm{Tr}$ stands for the trace over the Nambu indices, $\Omega$ is the volume of the system, $\tau \in (0, \beta)$ is imaginary time and $\beta = 1/T$, where $T$ is the temperature (we set $\hbar=k_B=1$).
The Nambu field operators $\Psi_{\bm k}(\tau)$ are in the Heisenberg representation.

\textit{Short-range interaction.}
If we take into account only the short-range attractive interaction $U_0$, see Fig.~\ref{fig:1}(b), the static pair susceptibility, $\chi_{C}(T, \Delta)$, is represented by the standard Cooper ladder series in the leading logarithmic order, hence the subscript $C$~\cite{kulikPairSusceptibilityMode1981},
\begin{eqnarray}
	&& \hspace{-8pt} \chi_C (T, \Delta) = \frac{\chi_{0}(T, \Delta)}{1 + U_0 \chi_{0}(T, \Delta)} \, , \label{BCST} 
\end{eqnarray}
where $\chi_{0}(T, \Delta)$ is the pair susceptibility of the non-interacting electron gas.
The critical temperature $T_c$ and zero-temperature SC gap $\Delta$ follow from the poles of $\chi_C$: $U_0 \chi_0(T_c, 0) = -1$, $U_0 \chi_0 (0, \Delta) = -1$.
This results in the standard BCS prediction for $T_c = T_{\rm BCS}$ and $\Delta = \Delta_{\rm BCS}$~\cite{bardeen1957}, 
\begin{eqnarray}
	&& \frac{\Delta_{\rm BCS}}{T_{\rm BCS}} = \frac{\pi} {e^{C_{\mathrm{E}}}} \approx 1.76, \hspace{5pt} \Delta_{\rm BCS} = 2 \omega_0 e^{-\frac{1}{\gamma_0}} \, , \label{DeltaBCS} 
\end{eqnarray}
where $C_{\mathrm{E}} \approx 0.577$ is the Euler-Mascheroni constant, $\omega_0$ is the characteristic phonon frequency, $\gamma_0 = -U_0 N_F > 0$ is the dimensionless short-range coupling constant, and $N_F$ the density of states per spin at the Fermi energy. 
Here, we used $\chi_0(T, 0) = N_F \ln [2 e^{C_{\mathrm{E}}} \omega_0/(\pi T)]$ and $\chi_0(0, \Delta) = N_F \ln[2 \omega_0/\Delta]$.

\textit{Finite-range interaction.}
If the finite-range attractive interaction $U_f (\tau, \bm r)$ is taken into account, the Cooper ladder summation that lies at the heart of the BCS approximation is no longer justified. 
Instead, the pair-susceptibility can be evaluated by means of dimensional-reduction~\cite{miserevDimensionalReductionLuttingerWard2023a,hutchinsonSpinSusceptibilityInteracting2023}, and was computed for an arbitrary finite-range interaction in Ref.~\cite{miserevMicroscopicMechanismPair2024}. This susceptibility, $\chi_f(T)$, does not acquire a pole at finite $T=T_c$, but instead  exhibits a power-law singularity at $T \to 0$:
\begin{eqnarray}
&& \hspace{-15pt} \chi_f(T) = \frac{N_F}{\gamma_f} \left[\left(\frac{2 e^{C_{\mathrm{E}}} \Lambda}{\pi T}\right)^{\gamma_f} -1 \right], \label{chiT} \\
&& \hspace{-15pt} \gamma_f = -\frac{2}{\pi v_F} \int\limits_0^\infty dr \int\limits_{-\beta/2}^{\beta/2} d\tau \, U_f(\tau, r) \, , \label{gammaf}
\end{eqnarray}
where $1 \gg \gamma_f > 0$ is the dimensionless coupling constant for the finite-range interaction.
The cut-off $\Lambda$ is determined by the following equation,
\begin{eqnarray}
&& \hspace{-15pt} \ln \Lambda  =  \int\limits_{-\beta/2}^{\beta/2}  d\tau  \int\limits_0^\infty  dr \, \frac{U_f(\tau, r)}{\pi v_F \gamma_f} \ln \left(\frac{r^2}{v_F^2} + \tau^2\right) -C_{\mathrm{E}} \, . \label{Lambda}
\end{eqnarray}
As a SC gap $\Delta$ regularizes $\chi_f$ the same way as finite $T$, an analogous power-law singularity emerges in the zero-temperature pair susceptibility $\chi_f(\Delta)$,
\begin{eqnarray}
	&& \chi_f(\Delta) = \frac{N_F}{\gamma_f} \left[\left(\frac{2 \Lambda}{\Delta}\right)^{\gamma_f} -1 \right] \, . \label{chiDelta}
\end{eqnarray}
Details of the calculations are outlined in the Supplemental Material  (SM)~\cite{SM,loboUnexpectedBehaviourIR1995,castillo-lopezNearfieldRadiativeHeat2020,mincheongFirstPrincipleCalculation2024} and are based on results of Ref.~\cite{miserevMicroscopicMechanismPair2024}. 
We point out that long-range SC order is unstable at any $T > 0$ as $\chi_f(T)$ is finite and $\chi_f(\Delta)$ is finite at any $\Delta > 0$.
It has been shown in Ref.~\cite{miserevMicroscopicMechanismPair2024} that there are zero-sound contributions competing with the Cooper pairing channel that destabilize the long-range SC order if the interaction is of the forward-scattering (finite-range) type.
This constitutes a paradigmatic difference from the BCS theory, which predicts finite $T_c$ and $\Delta$ for any attractive interaction.

\begin{figure}[t]
	\centering
	\includegraphics[width=0.99\columnwidth]{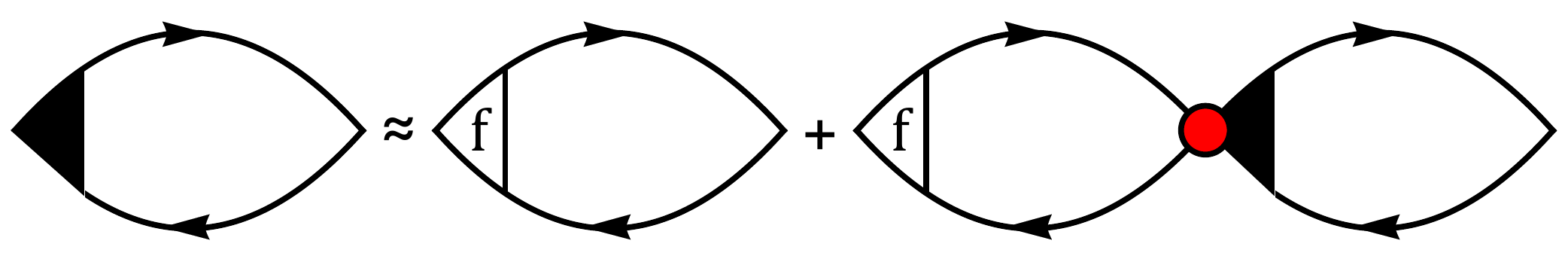}
	\caption{The full pair susceptibility $\chi$ [see Eq.~(\ref{fullchi})], represented by the bubble with the black triangle, is approximated by the Cooper ladder where each pair vertex is dressed by the finite-range interaction $U_f(i \omega, q)$ (white triangle denoted by ``f'').
	Red circle stands for $-U_0$, solid black lines stand for the  Nambu Green function $G(i\omega, \bm k) = (i \omega - H_0)^{-1}$, where $H_0$ is the single-particle mean-field Hamiltonian.
	The bubble with  triangle f represents $\chi_f(T,\Delta)$, see Eqs.~(\ref{chiT}) and (\ref{chiDelta}).
	In the case $U_f(i \omega, q) = 0$, this equation represents the standard Cooper ladder yielding the BCS susceptibility $\chi_C$ given by Eq.~(\ref{BCST}).}
	\label{fig:2}
\end{figure}

\textit{Combined interaction.}
Finally, we consider the combined effect of both finite-range and contact interactions.
For this, we dress the Cooper ladder diagrams by the finite-range interaction as shown in Fig.~\ref{fig:2},
\begin{eqnarray}
	&& \chi(T, \Delta) = \frac{\chi_f (T, \Delta)}{1 + U_0 \chi_f(T, \Delta)} \, , \label{fullchi}
\end{eqnarray}
where $\chi_f(T, \Delta)$ corresponds to $\chi_f(T)$ at $\Delta = 0$, see Eq.~(\ref{chiT}), and to $\chi_f(\Delta)$ at $T = 0$, see Eq.~(\ref{chiDelta}).
We emphasize that only the zero-momentum pair vertex is enhanced by an attractive finite-range interaction~\cite{miserevMicroscopicMechanismPair2024}.
On the contrary, the charge vertex remains unaffected by the attractive interaction, except for the addition of an irrelevant non-analyticity at $|q - 2k_F| \lesssim 1/R_s \ll k_F$.
This justifies the approximation shown in Fig.~\ref{fig:2} up to leading order in $\chi_f(T, \Delta)$.

The power-law singularity of $\chi_f(T, \Delta)$ is much stronger than the logarithmic singularity of $\chi_0(T, \Delta)$. This is due to enhanced range of the pair fluctuations, $\chi_f(r) \propto 1/r^{D - \gamma_f}$ vs. $\chi_0(r) \propto 1/r^D$, where $\chi_f(r)$ ($\chi_0(r)$) is the static pair susceptibility of the electron gas with (without) an attractive finite-range interaction, see Ref.~\cite{miserevMicroscopicMechanismPair2024}.

The pole of the static pair susceptibility [see Eq.~(\ref{fullchi})] provides explicit expressions for the critical temperature, $T_c$, and the zero-temperature gap, $\Delta$,
\begin{eqnarray}
	&& \Delta = \frac{\pi}{e^{C_{\mathrm{E}}}} T_c = 2 \Lambda e^{- \frac{1}{\Gamma}} , \hspace{5pt}  \Gamma = \frac{\gamma_f}{\ln \left(1 + \frac{\gamma_f}{\gamma_0}\right)} \, , \label{Delta} 
\end{eqnarray}
where $\Gamma$ is an effective coupling constant. 
The BCS limit corresponds to $\gamma_f \to 0$ resulting in $\Gamma \to \gamma_0$.
The opposite limit when $\gamma_0 \to 0$ results in $\Gamma \to 0$ and $T_c \to 0$, i.e. there is no SC order if only the finite-range interaction is present.

The ratio $\Delta/T_c \approx 1.76$ remains the same as in BCS theory, even though the BCS approximation is no longer valid in the presence of the finite-range interaction.
This is due to the limit $\gamma_f \ll 1$, which corresponds to the resummation of the leading logarithmic contributions.
In this limit, the first-order Cooper diagram determines the cut-off, see SM~\cite{SM}.

\begin{figure}[t]
	\centering
	\includegraphics[width=\columnwidth]{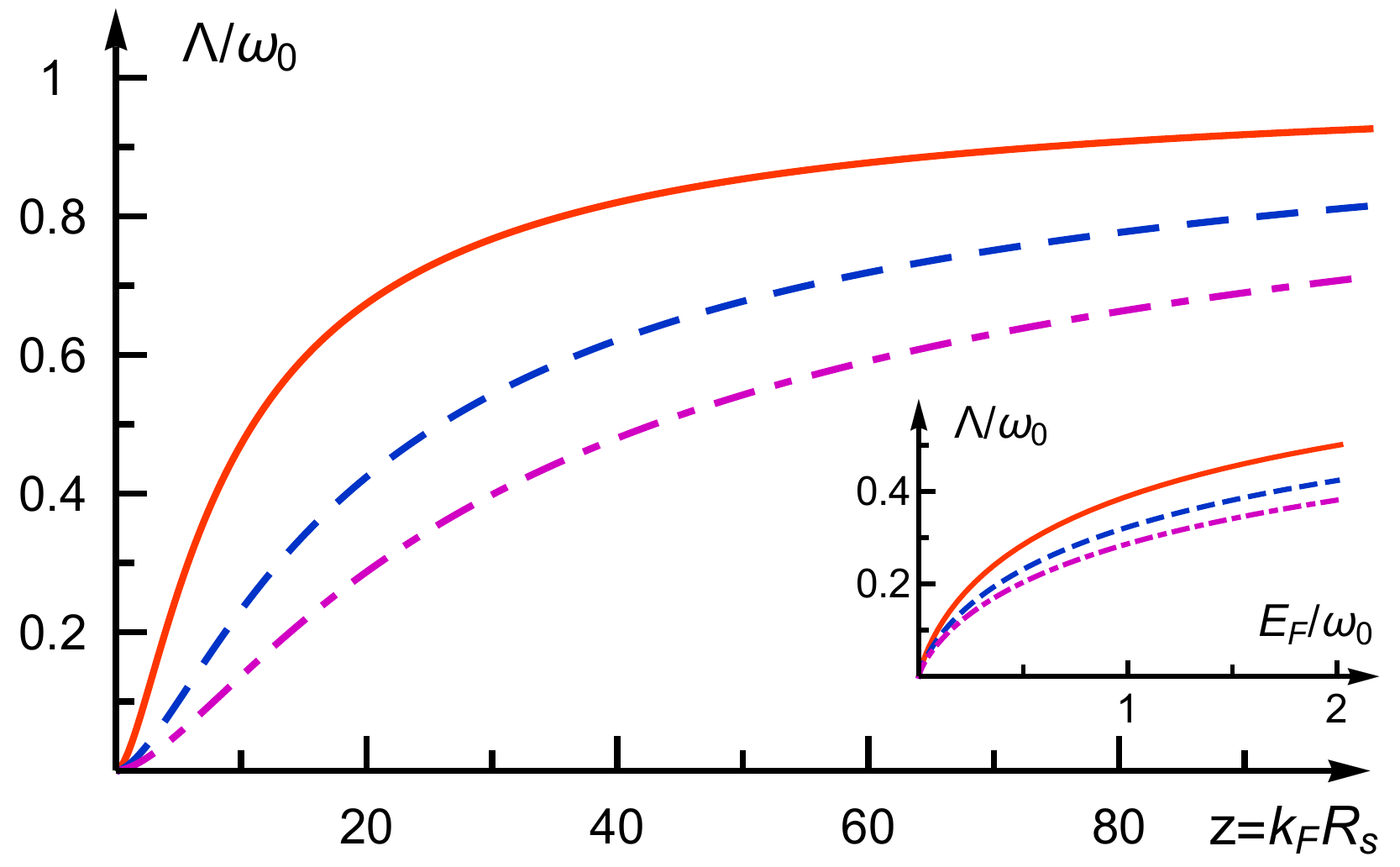}
	\caption{The cut-off $\Lambda$, see Eq.~(\ref{Lambda}), corresponding to the interaction in Eq.~(\ref{D2D}) at $\tilde{\omega}_0 = m R_s^2 \omega_0 = 30$ (red solid line), $\tilde{\omega}_0 = 100$ (blue dashed line), $\tilde{\omega}_0 = 200$ (violet dash-dotted line) as a function of dimensionless Fermi momentum $z = k_F R_s$. Inset: non-linear dependence of $\Lambda/\omega_0$ vs. $E_F/\omega_0$ at $E_F \lesssim \omega_0$ when $\Lambda \propto \sqrt{E_F v_F/R_s}$.}
	\label{fig:3}
\end{figure}

\textit{EPI model.}
As an example, we apply our theory to strongly anisotropic 3D materials where electrons are constrained to move within weakly coupled 2D crystal planes.
We assume that electrons interact with the bulk longitudinal optical (LO) phonons polarized across the crystal planes,
\begin{eqnarray}
	&& \hspace{-15pt} N_F U (i \omega, \bm q) = -\frac{\lambda_{ep}}{\sqrt{(q R_{s})^2 + 1}} \frac{\omega_0^2}{\omega^2 + \omega_0^2} \, , \label{D2D}
\end{eqnarray}
where $U(i \omega, \bm q)$ is an attractive electron-electron interaction mediated by phonons, $N_F = m/(2 \pi)$ the 2D density of states per spin, $m$ the 
effective mass, $\omega$ the bosonic Matsubara frequency, $\bm q$ the in-plane phonon momentum (the out-of-plane phonon momentum is integrated out), $q = |\bm q|$, $\omega_0$ the LO phonon frequency, $R_s$ the Coulomb screening length, $\lambda_{ep} = g_{ep}^2 m a_\parallel^2/(\pi \omega_0)$ the dimensionless coupling constant, $a_\parallel$ the in-plane lattice constant, $g_{ep}$ the EPI coupling at $\bm q = 0$. 
A detailed derivation of Eq.~\eqref{D2D} is provided in the SM~\cite{shamElectronPhononInteraction1963,bruus2004,SM}.
This model is relevant to cuprate materials where electrons are mobile within weakly coupled copper oxide planes, and interact with the buckling phonons polarized perpendicular to the planes \cite{devereaux1995,jepsen1998,opelPhysicalOriginBuckling1999,devereaux2004,cukCoupling$B_1g$Phonon2004,johnston2010}.

$U(i \omega, \bm q)$ is peaked at $q \lesssim 1/R_s$ as shown in Fig.~\ref{fig:1}(b).
The finite-range part of $U(i \omega, \bm q)$ is chosen such that $U_f(i \omega, 2k_F) = 0$ at all frequencies $\omega$, hence $U(i \omega, 2 k_F)$ is attributed to the short-range interaction.
The coupling constants $\gamma_0 = -U_0 N_F$ and $\gamma_f$ [see Eq.~(\ref{gammaf})], are then given by
\begin{eqnarray}
	&& \hspace{-10pt} \gamma_f (z) = \frac{2 \lambda_{ep}}{\pi z} \left(\mathrm{arcsinh}(2 z) - \frac{2z}{\sqrt{(2 z)^2 + 1}}\right) \, , \label{gammafz} \\ [3pt]
	&& \hspace{-10pt} \gamma_0 (z) = \frac{\lambda_{ep}}{\sqrt{(2 z)^2 + 1}} \, , \label{gamma0z} 
\end{eqnarray}
where $z = k_F R_{s}$ is the dimensionless Fermi momentum and $U_0 = U(0, 2 k_F)$.

The cut-off $\Lambda$ follows from Eq.~(\ref{Lambda}).
An explicit expression for $\Lambda$ and its asymptotic behavior is provided in the SM~\cite{SM}.
Here we plot it as function of $z$ at different values of the dimensionless phonon frequency $\tilde{\omega}_0 = m R_s^2 \omega_0$ in Fig.~\ref{fig:3}.
At large densities when $z \gg \tilde{\omega}_0$ or $v_F \gg R_s \omega_0$, $\Lambda \approx \omega_0$, i.e. the cut-off is similar to the BCS one, see Eq.~(\ref{DeltaBCS}). 
In the opposite limit of small densities, $\Lambda \propto \sqrt{E_F v_F/R_s} \ll E_F$ depends strongly on the electron density and does not depend on $\omega_0$, see the inset of Fig.~\ref{fig:3}. 
In order to access the low-density regime while maintaining the forward-scattering condition $z \gg 1$, we consider $\tilde{\omega}_0 \gg 1$.

\begin{figure}[t]
	\centering
	\includegraphics[width=\columnwidth]{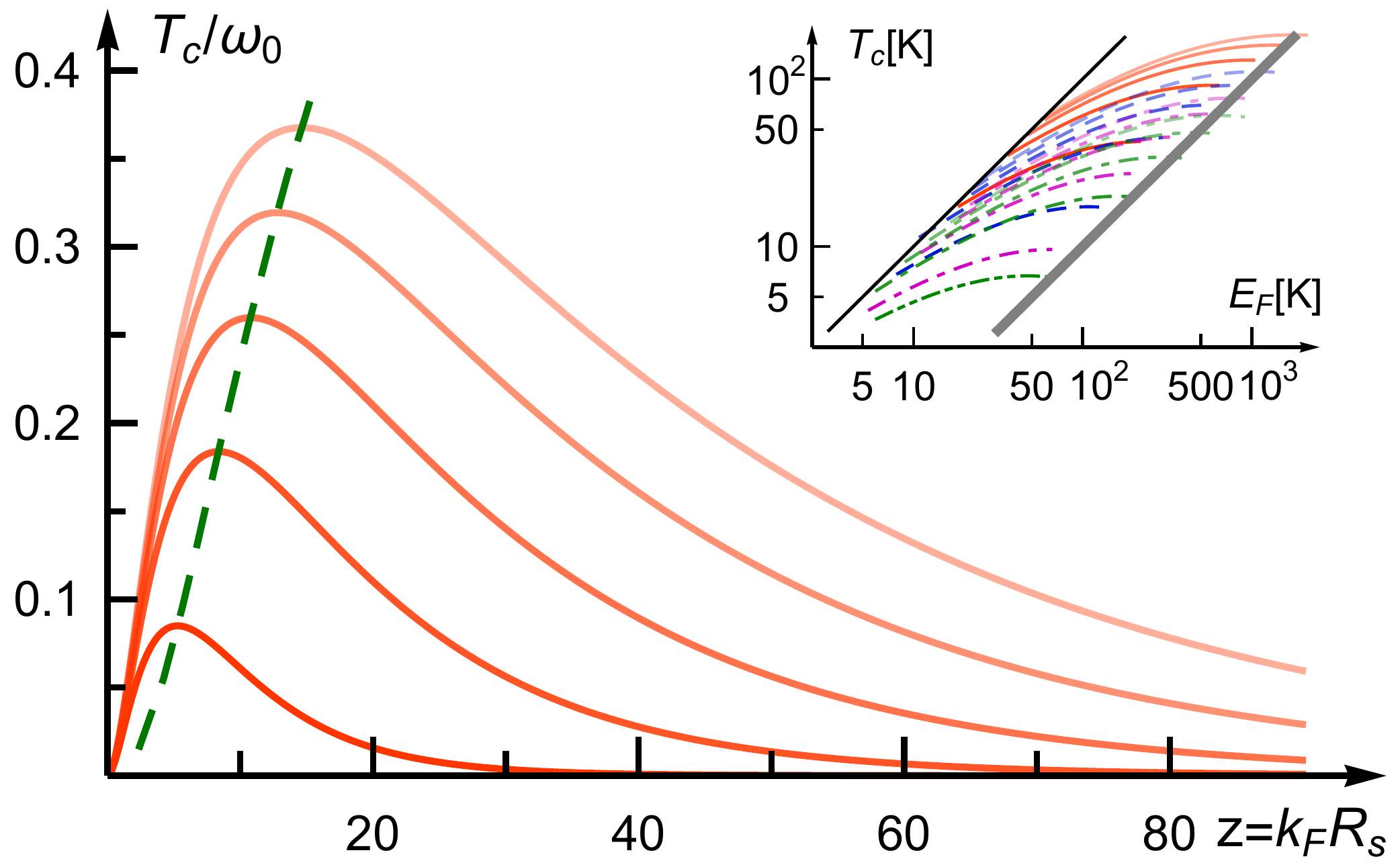}
	\caption{Critical temperature $T_c$ [see Eq.~(\ref{Delta})] in units of $\omega_0$ at $\tilde{\omega}_0 = m R_s^2 \omega_0 = 30$ as a function of dimensionless Fermi momentum $z = k_F R_s$ at different values of $\lambda_{ep} = 4, 8, 12, 16, 20$ shown in order of increasing line transparency. 
	The green dashed curve corresponds to the maximal $T_c^*$ vs. the optimal doping $z^*$. Inset: the Uemura plot $T_c$ vs. $E_F$ at $z < z^*$ for $\tilde{\omega}_0 = 30$ (red solid lines), $100$ (blue dashed lines), $200$ (violet dash-dotted lines), $300$ (green dash-dot-dotted lines), different values of $\lambda_{ep} \in [4, 20]$ are shown in order of increasing line transparency. We also choose $\omega_0 = 43\,$meV, another choice of $\omega_0$ does not change the landscape of curves confined between $T_c = E_F$ (black solid line) and $T_c = 0.1 E_F \sim T_c^*$ (thick gray line).}
	\label{fig:4}
\end{figure}

\textit{Superconducting dome.}
We plot $T_c$ [see Eq.~(\ref{Delta})] versus dimensionless momentum $z = k_F R_s$ for different values of $\lambda_{ep}$ at $\tilde{\omega}_0 = 30$ in Fig.~\ref{fig:4}.
$T_c$ features a dome structure with a clear maximum $T_c^*$ at the optimal doping corresponding to $k_F^* \gg 1/R_s$.
The physical reason behind the dome structure can be understood as follows. 
The coupling constant $\lambda(z) \to 0$ at large densities corresponding to $z \gg \tilde{\omega}_0$ [see Eqs.~(\ref{Delta}), (\ref{gammafz}), (\ref{gamma0z})], which results in an exponentially small gap.
The cutoff stemming from the finite-range interaction tends to zero at small densities, $\Lambda \propto \sqrt{E_F v_F/R_s}$.
Therefore, there is a maximal $T_c^*$ corresponding to the optimal $k_F^*$ where the two behaviors compete.
We also plot $T_c$ vs. $E_F$ in the underdoped regime when $k_F < k_F^{*}$ at many different values of $\tilde{\omega}_0$ and $\lambda_{ep}$, see the inset of Fig.~\ref{fig:4}, and observe that $T_c^*/E_F \sim 0.1$ (thick gray line) for the entire landscape of curves which is consistent with observations in cuprates \cite{uemuraUniversalCorrelations1989,uemuraClassifyingSuperconductorsPlot1991,takenakaStronglyCorrelatedSuperconductivity2021}.

The SC dome has been also predicted for isotropic 3D metals with toy models for finite-range interactions treated within the BCS theory \cite{langmann2019}.
Here, we show that the BCS theory breaks down for interactions with finite-range $R_s \gg \lambda_F$, and that the residual short-range interaction plays a crucial role in stabilizing the SC order.
Moreover, Eq.~(\ref{Lambda}) for the cut-off takes into account the retardation effect resulting in $\Lambda \ll \omega_0$ at   $\omega_0 \gg v_F/R_s$.
Next, we show that such a behavior of the cut-off results in a strong reduction of the isotope effect.

\begin{figure}[t]
	\centering
	\includegraphics[width=\columnwidth]{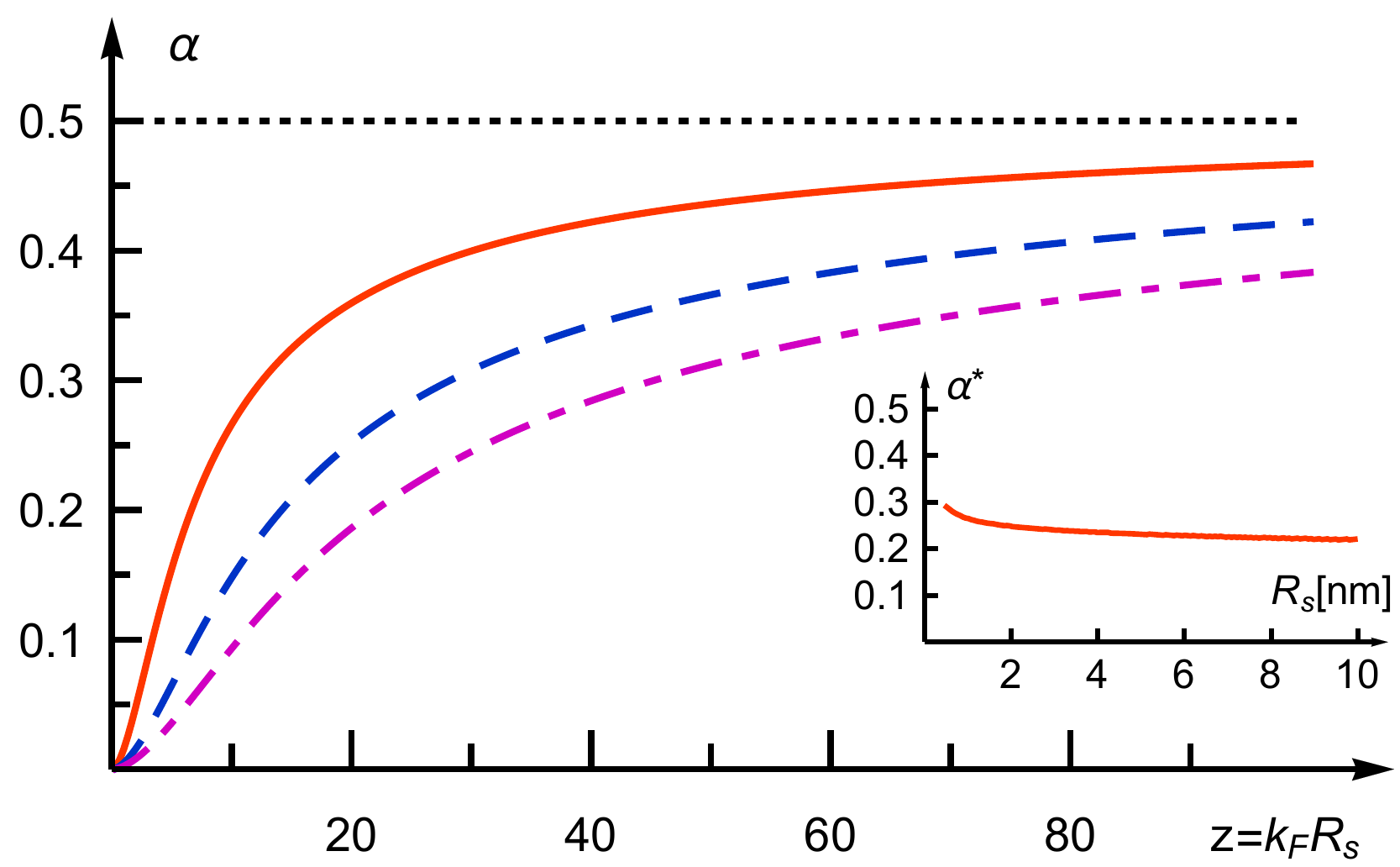} 	
	\caption{Isotope exponent $\alpha$ [see Eq.~(\ref{alpha})] vs. dimensionless momentum $z= k_F R_s$ at $\tilde{\omega}_0 = m R_s^2 \omega_0 = 30$ (red solid line), $\tilde{\omega}_0 = 100$ (blue dashed line), $\tilde{\omega}_0 = 200$ (violet dash-dotted line). Black dotted line corresponds to $\alpha_{\mathrm{BCS}}=0.5$. Inset: $\alpha^*$ vs. $R_s$ corresponding to YBCO parameters at the optimal doping.}
	\label{fig:5}
\end{figure}

\begin{figure}[t]
	\centering
	\includegraphics[width=0.49\columnwidth]{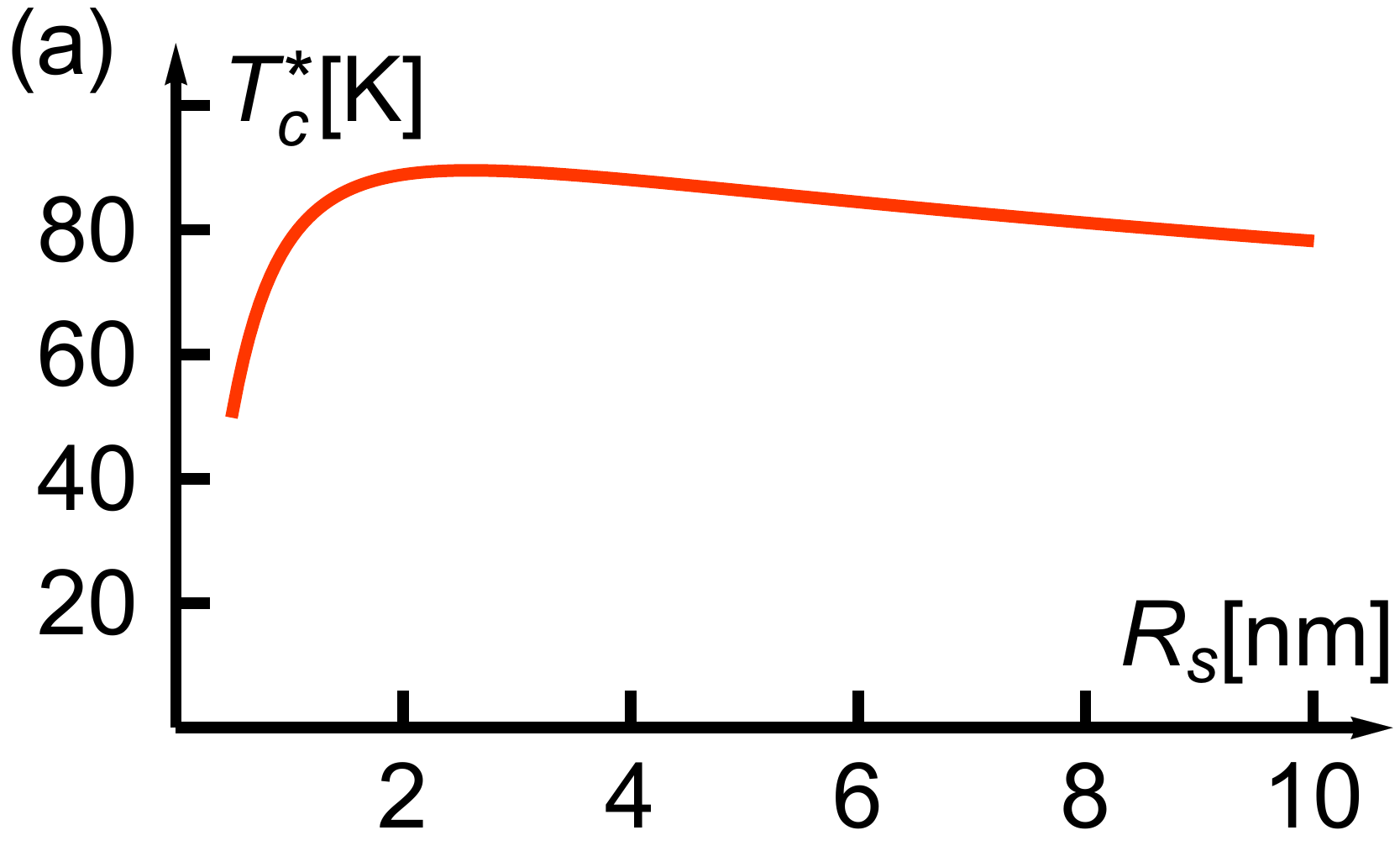} 	\includegraphics[width=0.49\columnwidth]{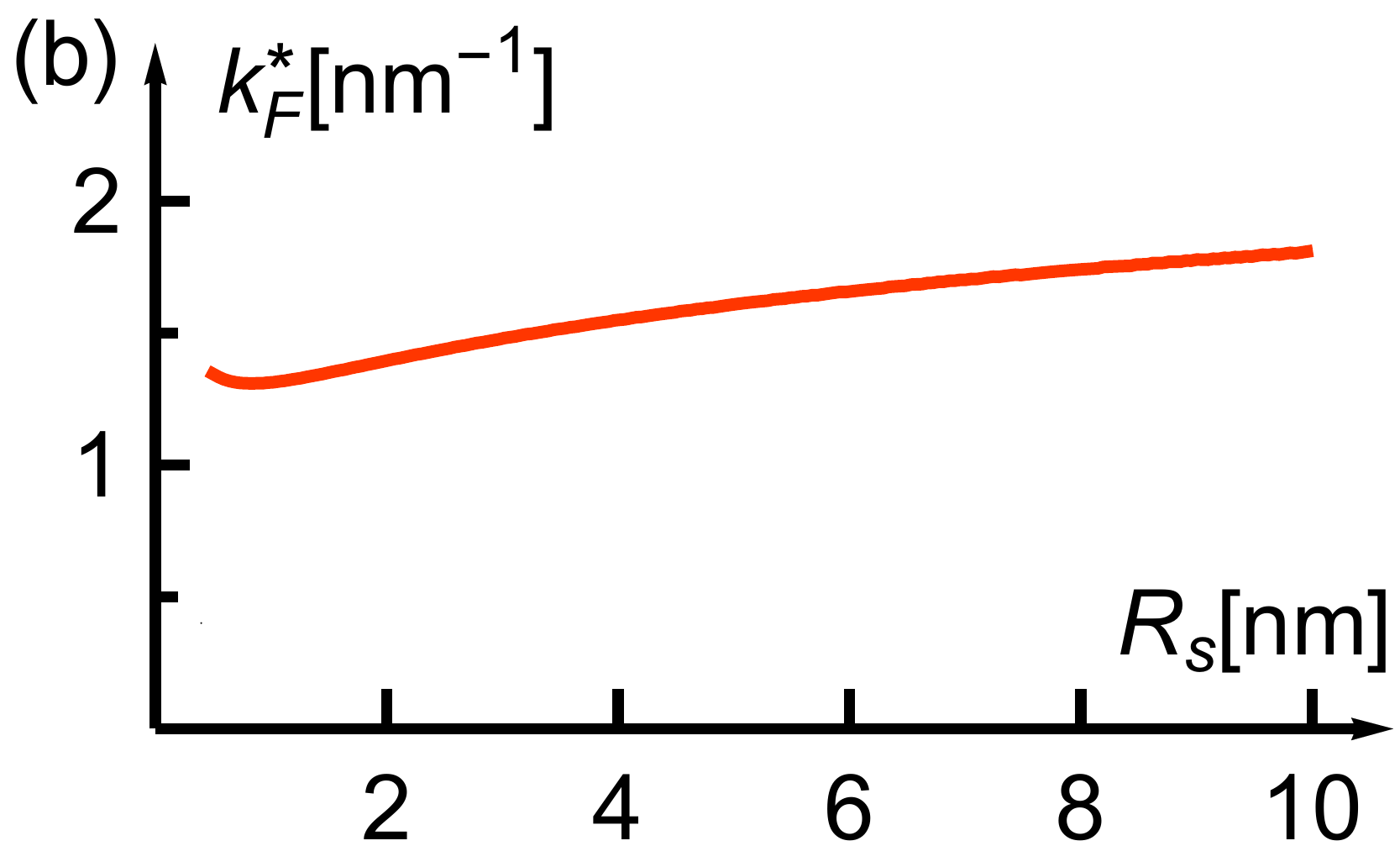}
	\caption{(a) $T_c^{*}$ vs.~$R_s$ and (b) $k_F^{*}$ vs.~$R_s$ plotted for YBCO parameters. At $R_s \gtrsim 1\,$nm, $T_c^{*} \approx 80\,$K and $k_F^{*}\approx 1.5\,$nm$^{-1}$. }
	\label{fig:6}
\end{figure}

\textit{Isotope effect}.
The isotope effect that is described by the exponent $\alpha$,
\begin{eqnarray}
	&& \alpha = -\frac{d \ln T_c}{d \ln M} = \frac{1}{2} \frac{d \ln \Lambda}{d \ln \omega_0} \, , \label{alpha}
\end{eqnarray}
where $M$ is the ion mass, and $\omega_0 \propto 1/\sqrt{M}$.
Here, we point out that neither $\lambda_{ep}$ nor $\Gamma$ depend on the ion mass $M$, so $T_c$ depends on $M$ only through the cut-off $\Lambda$ (see Fig.~\ref{fig:3}). The BCS prediction $\alpha_{\mathrm{BCS}} = 0.5$ stems from Eq.~(\ref{DeltaBCS}) where $\Lambda = \omega_0$.
We plot $\alpha$ versus the dimensionless Fermi momentum $z = k_F R_s$ at different values of $\tilde{\omega}_0 = 30, 100, 200$ in Fig.~\ref{fig:5} where we observe a strong reduction of $\alpha$ at $z \ll \tilde{\omega}_0$ ($v_F \ll R_s \omega_0$) compared to its BCS value.

\textit{Dependence of $T_c^*$, $k_F^*$, and $\alpha^*$ on $R_s$.}
The interaction range $R_s$ is considered as a phenomenological parameter within our model.
Reliable estimates of $R_s$ in metals can be obtained using the speed of longitudinal acoustic phonons, $c_{\mathrm{ac}} \approx \omega_{ip} R_s$, where $\omega_{ip}$ is the ionic plasma frequency \cite{ashcroft1976,mahanManyParticlePhysics2000}.
For example, $\omega_{ip} \approx 3\,$meV, $c_{\mathrm{ac}} \approx 5 \times 10^3\,$m/s \cite{elbaumAcousticStudiesHigh1996} in YBCO near the optimal doping yielding $R_s \approx 2\,$nm (see the SM~\cite{SM}).

We plot $T_c^{*}$ and $k_F^{*}$ in Fig.~\ref{fig:6} for YBCO parameters: $\omega_0 = 43\,$meV, $m = 2 m_0$~\cite{padillaConstantEffectiveMass2005}, $g_{ep} \approx 0.15\,$eV.
Here, we take a median EPI over known estimates, $g_{ep} \in (0.1, 0.2)  \,$eV~\cite{devereaux1995,jepsen1998,opelPhysicalOriginBuckling1999,devereaux2004,cukCoupling$B_1g$Phonon2004,johnston2010}.
These estimates are performed for a short-range EPI with $R_s \sim 0.5 \,$nm \cite{johnston2010}. Taking into account that $\lambda_{ep} \propto R_s$ (see SM~\cite{SM}), we find $\lambda_{ep} \approx 1.5 R_s[\mathrm{nm}]$ and $\tilde{\omega}_0 \approx 1.2 R_s^2 [\mathrm{nm}^2]$.
At $R_s \gtrsim 1\,$nm we find $T_c^* \approx 80\,$K and $k_F^* \approx 1.5\,$nm$^{-1}$. 
We point out that $k_F^* = \sqrt{2 \pi p^*/a_\parallel^2} \approx 2.6 \,$nm$^{-1}$ in YBCO, where $p^* = 0.16$ holes per Cu atom, $a_\parallel = 3.8 \, \text{\AA}$.
We also note that the rapid reduction of $T_c^*$ at $R_s < 2\,$nm and its near-constant value at $R_s > 2\,$nm, see Fig.~\ref{fig:6}(a), shows a similar trend observed in a recent experiment in the proximity-screened magic-angle twisted bilayer graphene \cite{barrierCoulombScreeningSuperconductivity2024}.

Interestingly, $T_c^*$ and $k_F^*$ calculated for the s-wave SC using the YBCO parameters agree well with the experimental data.
However, we emphasize that rigorous SC calculations corresponding to high-$T_c$ materials should be performed for a $d$-wave SC order parameter originating from $d$-wave EPI \cite{devereaux1995,jepsen1998,opelPhysicalOriginBuckling1999,devereaux2004,cukCoupling$B_1g$Phonon2004,johnston2010}.
The generalization of our theory to the $d$-wave SC order is technically straightforward yet somewhat cumbersome.
As our main goal here is to demonstrate new features of our theory, such generalizations are left for future work.

The inset of Fig.~\ref{fig:5} shows the isotope exponent $\alpha^*$ at the optimal doping versus $R_s$ calculated for estimated YBCO parameters. 
We find that $\alpha^* \approx 0.25$ which is a factor of $2$ smaller than the BCS value $\alpha_{\mathrm{BCS}} = 0.5$.
The experimental value of $\alpha^* \approx 0.1$ in YBCO is still a factor of $2$ smaller than our prediction which presumably can be attributed to the $d$-wave nature of SC in cuprates~\cite{crawfordOxygenIsotopeEffect1990,franckCopperOxygenIsotope1993,pringleEffectDopingImpurities2000,zhaoUnconventionalIsotopeEffects2001}.

\textit{Conclusion.} Remarkably, our model with finite-range interaction captures several SC phenomena that cannot be reconciled with BCS theory, including a strongly reduced isotope effect, SC dome, and large values of both the optimal doping and $T_c^*$. These phenomena have been the cause of much attention paid to unconventional pairing mechanisms. In this Letter, we demonstrated that a finite-range EPI in layered materials produces many of the key features that are commonly observed in high-$T_c$ SCs, heavy-fermion materials, magic-angle twisted bilayer graphene, and other quantum materials. We stress that these SC phenomena are strongly non-BCS-like as the BCS approximation breaks down for any finite-range interaction.

{\it Acknowledgments}. This work was supported by the Georg H. Endress Foundation and the Swiss National Science Foundation (SNSF).

\section*{Supplementary Material for ``High-Temperature Superconductivity from Finite-Range Attractive Interaction''}

Here, we derive the cut-off $\Lambda$, the coupling constant $\gamma_f$, the electron-phonon interaction (EPI) $U(i \omega, \bm q)$, and provide an estimate for the screening length $R_s$.

\begin{widetext}
	
	\begin{figure}[h!]
		\centering
		\includegraphics[width=0.99\textwidth]{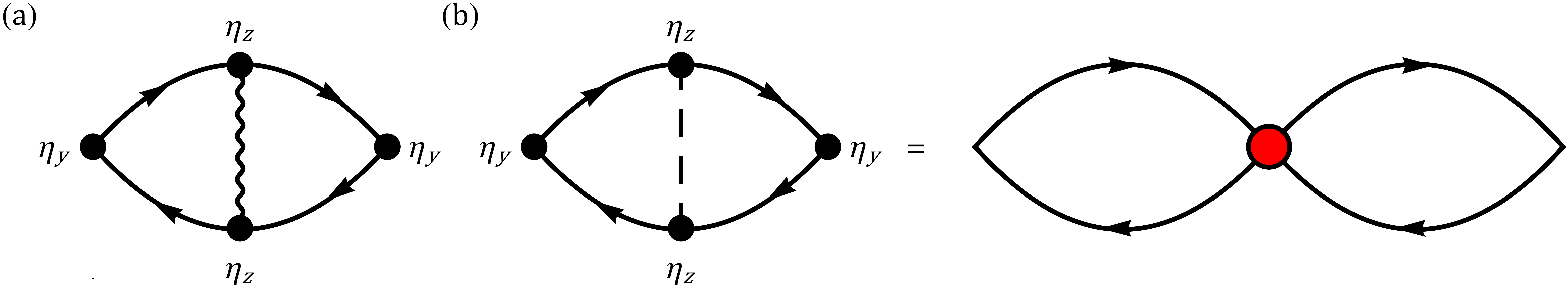}
		\caption{(a) Leading first-order correction to the pair susceptibility, see Eq.~(\ref{Dchi}), \cite{miserevDimensionalReductionLuttingerWard2023a,hutchinsonSpinSusceptibilityInteracting2023,miserevMicroscopicMechanismPair2024}. Solid lines correspond to the Green function [see Eq.~(\ref{G})], wavy line stands for the finite-range interaction $U_f(i \omega, \bm q)$.
		(b) Similar diagram but with the short-range interaction $U_0$ (dashed line). In the main text, we use diagrams with collapsed interaction vertices where the dashed line is represented by the red circle (standing for $-U_0$). Nevertheless, the trace must be taken the same way as in the diagram on the left-hand side including all $\eta_i$ matrices.}
		\label{figSM:1}
	\end{figure}
	
\end{widetext}

We consider an interacting electron system described by the following Euclidean action,
\begin{eqnarray}
	&& \hspace{-10pt} \mathcal{S} = \int dz \, \overline{\Psi}(z) \left(\partial_{\tau} + H_0\right) \Psi(z) \nonumber \\
	&&  + \frac{1}{2} \int dz dz' \, U(z-z') \varrho(z) \varrho(z') \, ,
\end{eqnarray}
where $z = (\tau, \bm r)$, $\tau \in (0, \beta)$ is the imaginary time, $\beta = 1/T$, $T$ the temperature, $\bm r$ the $D$-dimensional spatial coordinate, $\Psi(z)$ ($\overline{\Psi}(z)$) the fermion (conjugate) field operator in the Nambu representation, $H_0$ the single-particle mean-field Hamiltonian, $U(z)$ the interaction, $\varrho(z) = \overline{\Psi}(z) \eta_z \Psi(z)$ the particle density operator, $\eta_z$ the $2\times 2$ Nambu Pauli matrix.

\section{The cut-off $\Lambda$ and effective forward-scattering coupling $\gamma_f$}
\label{sec:cutoff}

Here we derive the cut-off $\Lambda$ and the forward-scattering coupling $\gamma_f$ corresponding to an arbitrary finite-range interaction $U_f(\tau, \bm r)$.
For this, we consider the leading first-order correction to the $D$-dimensional pair susceptibility, $\chi^{(1)}(\tau, r)$, shown in Fig.~\ref{figSM:1}(a).
After dimensional reduction~(see Refs.~\cite{miserevDimensionalReductionLuttingerWard2023a,hutchinsonSpinSusceptibilityInteracting2023,miserevMicroscopicMechanismPair2024}), 
which strongly simplifies the evaluation of the diagrams,
it takes the following form, 
\begin{eqnarray}
	&& \hspace{-15pt} \chi^{(1)}(\tau, r) = \frac{2 \chi_{1D}^{(1)} (\tau, r)}{(\lambda_F r)^{D - 1}} \, , \label{Dchi}\\
	&& \hspace{-15pt} \chi_{1D}^{(1)}(\xi) = \frac{1}{2} \int d\xi_1 d\xi_2 \, U_f(\xi_1 - \xi_2) \nonumber \\
	&& \hspace{5pt} \times \mathrm{Tr}\left[ \eta_y g(\xi - \xi_1) \eta_z g(\xi_1)  \eta_y g(-\xi_2) \eta_z g(\xi_2 - \xi) \right] \, , \label{chiP1D} 
\end{eqnarray}
where $\xi = (\tau, x)$, $x \in (-\infty, \infty)$ is an effective 1D coordinate, $\tau \in (0, \beta)$ is imaginary time, $d \xi_i = dx_i d\tau_i$, $U_f(\tau, x) = U_f(\tau, |x|)$ is the finite-range interaction taken at $r = |x|$, $\eta_z$ corresponds to the charge vertex in the Nambu basis, $\eta_y \propto \delta H_0/\delta \phi_{\mathrm{SC}}$, $\phi_{\mathrm{SC}}$ is the SC phase, $\mathrm{Tr}$ stands for the trace over the Nambu indices and the factor $1/2$ is introduced to normalize the trace (see Eq.~(1) in the main text). Here, $g(\xi)$ is the effective 1D Green function,
\begin{eqnarray}
	&& \hspace{-15pt} g (\tau, x) = T \sum\limits_{\omega_n} \int\limits_{-\infty}^\infty \frac{dq}{2 \pi} e^{i q x - i \omega_n \tau} G(i \omega_n, q) \, , \label{g1D} \\
	&& \hspace{-15pt} G(i \omega_n, q) = (i \omega_n - H_0)^{-1} = - \frac{i \omega_n + q v_F \eta_z + \Delta \eta_x}{\omega_n^2 + (q v_F)^2 + \Delta^2} \, , \label{G}
\end{eqnarray}
where $H_0 = q v_F \eta_z + \Delta \eta_x$ is the mean-field Hamiltonian, $q = k - k_F$, $k$ is the absolute value of the electron momentum $\bm k$, $\Delta$ is the $s$-wave SC gap, and
$\omega_n = \pi T (2 n + 1)$ is the fermionic Matsubara frequency with $n$ being an integer.
Here, we use the Nambu basis $\Psi_{\bm k}^T = (c_{\bm k, \uparrow}, c_{-\bm k, \downarrow}^\dagger)$, where $c_{\bm k, \sigma}$ ($c_{\bm k, \sigma}^\dagger$) corresponds to the annihilation (creation) operator with momentum $\bm k$ and spin $\sigma \in \{\uparrow, \downarrow\}$.
The zero-frequency and zero-momentum pair susceptibility then follows from Eq.~(\ref{Dchi}),
\begin{eqnarray}
	&& \hspace{-15pt} \chi^{(1)} = \int \chi^{(1)}(\tau, r) \, d\tau d \bm r = 2 \pi v_F N_F \int \chi^{(1)}_{1D}(\xi) \, d\xi \, , \label{Chistatic}
\end{eqnarray}
where $N_F$ is the density of states at the Fermi level per band.
It is convenient to introduce the following matrix,
\begin{eqnarray}
	&& F(\xi) = \int g(\xi') \eta_y g(\xi - \xi') \, d\xi' \, . \label{Fxi}
\end{eqnarray}
Then, the first-order susceptibility takes the following form,
\begin{eqnarray}
	&& \hspace{-15pt} \chi^{(1)} = 2 \pi v_F N_F \int d\xi \, U_f (\xi) \frac{1}{2} \mathrm{Tr}\left[F(-\xi) \eta_z F(\xi) \eta_z\right] \, . \label{ChiF}
\end{eqnarray}
Substituting Eq.~(\ref{g1D}) into Eq.~(\ref{Fxi}), we find,
\begin{eqnarray}
	&& F(\xi) = -f(\xi) \eta_y \, , \\
	&& f(\xi) = T \sum\limits_{\omega_n} \int\limits_{-\infty}^\infty \frac{d q}{2 \pi} \frac{e^{i q x - i \omega_n \tau}}{\omega_n^2 + (q v_F)^2 + \Delta^2} \, . \label{f}
\end{eqnarray}
Therefore, the first-order correction to the pair susceptibility due to finite-range interaction can be represented as follows,
\begin{eqnarray}
	&& \chi^{(1)} = - 2\pi v_F N_F \int d\xi \, U_f(\xi) f(\xi) f(-\xi) \, . \label{chif2}
\end{eqnarray}

At $T = 0$, $f(\xi)$ corresponds to the modified Bessel function of the second kind, $K_0$,
\begin{eqnarray}
	&& f(\xi, T = 0) = \frac{K_0(\Delta R)}{2 \pi v_F} \, , 
\end{eqnarray}
where $R = \sqrt{\tau^2 +x^2/v_F^2}$.
Then, the susceptibility correction takes the form,
\begin{eqnarray}
	&& \chi^{(1)}(\Delta) = -\frac{N_F}{2 \pi v_F} \int d\xi \, U_f(\xi) K_0^2(\Delta R) \, . \label{chi1K0}
\end{eqnarray}
The integral converges at $|x| \lesssim R_s$ and $\tau \lesssim 1/\omega_0$, where $\omega_0$ is the phonon frequency.
If $\Delta \ll \mathrm{min}\{\omega_0, v_F/R_s\}$, then $R\Delta  \ll 1$~and we can use the asymptotics of the Bessel function,
\begin{eqnarray}
	&& K_0(z) \approx \ln 2 - C_{\mathrm{E}} - \ln (z) \, ,
\end{eqnarray}
where $C_{\mathrm{E}}$ is the Euler-Mascheroni constant.
Substituting this asymptotic expression into Eq.~(\ref{chi1K0}), we find the expansion of $\chi^{(1)}$ in powers of $\ln \Delta$.
We have to compare this expansion with the one-loop RG result from Ref.~\cite{miserevMicroscopicMechanismPair2024} shown in Eq.~(7) in the main text,
\begin{eqnarray}
	&& \chi^{(1)} = \frac{\gamma_f N_F}{2} \ln^2\left(\frac{2 \Lambda}{\Delta}\right) \, .
\end{eqnarray}
Comparing coefficients next to the $\ln^2 (\Delta)$ terms provides the coupling constant,
\begin{eqnarray}
	&& \gamma_f = - \frac{1}{\pi v_F} \int d\xi \, U_f(\xi) \, , \label{gf}
\end{eqnarray}
where $d\xi = d\tau \, dx$ and we integrate over $x$ over the real line $x \in (-\infty, \infty)$.
This is equivalent to Eq.~(5) in the main text.
Comparing the coefficients next to $\ln \Delta$ provides the cut-off,
\begin{eqnarray}
	&& \ln \Lambda = \int d\xi \, \frac{U_f(\xi)}{\pi v_F \gamma_f} \ln \left(e^{C_{\mathrm{E}}} R\right) \, , \label{LambdaSM}
\end{eqnarray}
which is equivalent to Eq.~(6) in the main text. The remaining term is independent of $\Delta$ and can be omitted within the one-loop RG.

In the opposite limit, $\Delta = 0$ but finite $T$, we first sum over the fermionic frequencies in $f(\xi, \Delta=0)$ given in  Eq.~(\ref{f}), 
\begin{eqnarray}
	&& T \sum\limits_{\omega_n} \frac{e^{- i \omega_n \tau}}{\omega_n^2 + (q v_F)^2} = \frac{e^{-q v_F |\tau|}}{2 q v_F} \left(1 - n_F\left(q v_F\right)\right) \nonumber \\
	&& \hspace{83pt} - \frac{e^{q v_F |\tau|}}{{2} q v_F} n_F\left(q v_F\right) \, ,
\end{eqnarray}
where $\tau \in (-\beta, \beta)$
and $n_F(x)= (e^{\beta x} + 1)^{-1}$ is the Fermi-Dirac distribution function.
One way to evaluate this sum is to differentiate over $\tau$ twice and solve the resulting simple second-order differential equation with the antiperiodic condition that the sum changes sign at $\tau \to \tau + \beta$.
In order to take the remaining integral over $q$, we first differentiate $f(\xi, \Delta=0)$ with respect to $x$,
\begin{eqnarray}
	&& \hspace{-27pt} \frac{\partial f(\xi)}{\partial x} = -\frac{1}{2 \pi v_F^2} \mathrm{Im} \left\{\frac{1}{2} \int\limits_{-\infty}^\infty dQ \, e^{i Q \frac{x}{v_F}} \right. \nonumber \\
	&& \left. \times \left(\frac{e^{-Q |\tau|}}{e^{-\beta Q}+1}-\frac{e^{Q |\tau|}}{e^{\beta Q}+1}\right)  \right\} \nonumber \\
	&&  = -\frac{T}{2 v_F^2} \mathrm{Re} \left[\mathrm{csch}\left(\pi T \left(\frac{x}{v_F} + i \tau\right)\right)\right] \, , 
\end{eqnarray}
where $Q = q v_F$, and the integral over $Q$ was taken by expanding the denominators in a geometric series. 
Taking into account that $f(\xi) = 0$ at $x \to \pm \infty$, we find,
\begin{eqnarray}
	&& \hspace{-17pt} f(\xi, \Delta = 0) = \! \frac{1}{4 \pi v_F} \ln \! \left(\!\frac{\cosh\left(\pi T \frac{x}{v_F}\right) + \cos \left(\pi T \tau\right)}{\cosh\left(\pi T \frac{x}{v_F}\right) - \cos \left(\pi T \tau\right)}\!\right) \! .
\end{eqnarray}
As the integral in Eq.~(\ref{chif2}) converges at $|x| \lesssim R_s$ and $|\tau| \lesssim 1/\omega_0$, we can consider the asymptotics of $f(\xi)$ at $T \ll \mathrm{min}\{\omega_0, v_F/R_s\}$,
\begin{eqnarray}
	&& f(\xi, \Delta = 0) \approx - \frac{1}{2 \pi v_F} \ln\left(\frac{\pi T R}{2}\right) \, , 
\end{eqnarray}
where again $R = \sqrt{\tau^2 +x^2/v_F^2}$.
This gives the same asymptotics as $f(\xi, T = 0)$ with the substitution $\Delta \to \pi T/e^{C_{\mathrm{E}}}$.
This coincides with the expansion of the one-loop RG result given in Eq.~(4) in the main text.

We point out that the fact that the ratio $\Delta/T_c = \pi /e^{C_{\mathrm{E}}} \approx 1.76$ coincides with the BCS ratio is only due to the one-loop RG approximation and is expected to change with the inclusion of higher loops.
This coincidence does not justify the BCS approximation, which breaks down completely for any finite-range (forward scattering) interaction, see Ref.~\cite{miserevMicroscopicMechanismPair2024}.

We point out that a similar procedure can be used to derive the cut-off for the retarded short-range interaction, see Fig.~\ref{figSM:1}(b). 
For the short-range interaction, the interaction lines can be collapsed into a single vertex (red circle in Fig.~\ref{figSM:1}(b)). However, the trace over Nambu indices has to be taken over a single fermion loop, like in the original diagram. We use this representation in Fig.~2 in the main text.

\section{Electron-Phonon Interaction}

Here, we derive the effective EPI using standard textbook approach, see Refs.~\cite{shamElectronPhononInteraction1963,bruus2004},
\begin{eqnarray}
	&& H_{e-ph} = \int d^2 \bm r \, n(\bm r) \int d^3 \bm R \, \delta \rho(\bm R) V_C(|\bm R - \bm r|) , \label{eph} \\
	&& V_C(r) = -\frac{Z e^2}{\varepsilon_\infty r} \exp\left(-\frac{r}{R_s}\right) \, , \label{VC}
\end{eqnarray}
where $Z|e|$ is the ionic charge corresponding to a given phonon mode, $n(\bm r)$ is the density operator of electrons confined to a chosen 2D plane, $\delta \rho (\bm R)$ is the local positive charge imbalance due to vibrations of the 3D ion lattice, $V_C(r)$ is the 3D screened Coulomb interaction, $R_s$ is the 3D Coulomb screening length and $\varepsilon_\infty$ the high-frequency dielectric constant.

The charge imbalance is due to the longitudinal phonons slightly altering the unit cell size, 
\begin{eqnarray}
	&& \delta \rho (\bm R) = - \frac{\mathrm{div} \bm u (\bm R) }{\Omega_0} \, , \label{drho}
\end{eqnarray}
where $\Omega_0$ is the unit cell volume, and $\bm u(\bm R)$ is the displacement field of the ion lattice,
\begin{eqnarray}
	&& \hspace{-15pt} \mathrm{div} \bm u (\bm R) = \sum\limits_{\bm Q} \frac{\sqrt{\Omega_0} \left(\bm Q \cdot \bm e_{z}\right) e^{i \bm Q \cdot \bm R}}{\sqrt{2 \Omega M \omega_0}} i \left(b_{\bm Q} + b_{-\bm Q}^\dagger\right) \, , \label{divu}
\end{eqnarray}
where $\Omega$ is the full volume of the crystal, $M$ the ion mass, $\omega_0$ the phonon frequency, $\bm e_{z}$ the polarization perpendicular to the conducting planes, $b_{\bm Q}$ the annihilation operator of the longitudinal optical (LO) phonon with 3D momentum $\bm Q$. 
Therefore, the EPI can be represented as follows,
\begin{eqnarray}
	&& \hspace{-15pt} H_{e-ph} = \int d^2 \bm r \, n(\bm r) \phi (\bm r) \, , \label{Heph} \\
	&& \hspace{-15pt} \phi(\bm r) = -i \sum\limits_{\bm Q} \frac{Q_z V_C (\bm Q) e^{i \bm Q \cdot \bm r}}{\sqrt{2 \Omega \Omega_0 M \omega_0}} \left(b_{\bm Q} + b_{-\bm Q}^\dagger\right) \, , \label{phi}
\end{eqnarray}
where $\phi(\bm r)$ is the phonon field, and $V_C (\bm Q)$ is the 3D Fourier transform of the screened Coulomb interaction,
\begin{eqnarray}
	&& V_C (\bm Q) = - \frac{4 \pi Z e^2}{\varepsilon_\infty \left(Q^2 + R_s^{-2}\right)} \, . \label{VCQ}
\end{eqnarray}
Equations~\eqref{Heph} and~\eqref{phi} are just the Fourier transform of the standard electron-phonon interaction~\cite{bruus2004}.

Next, we integrate out the phonons using the 3D phonon propagator,
\begin{eqnarray}
	D(i \omega, \bm Q) &=& - \int d\tau \, \langle \mathcal{T} \left\{\phi(\tau, \bm Q) \phi(0, \bm Q)\right\} \rangle e^{i \omega \tau}  \nonumber\\
	&=& -\frac{1}{M \Omega_0} \frac{Q_z^2 |V_C (\bm Q)|^2}{\omega^2 + \omega_0^2} \, , \label{D3D}
\end{eqnarray}
where $\omega$ is the bosonic Matsubara frequency, and $\mathcal{T}$ is the time-ordering operator.

The effective interaction between electrons in the same 2D plane that is mediated by the LO phonon polarized perpendicular to the conducting planes is then the following,
\begin{eqnarray}
	&& \hspace{-39pt}	U (i \omega, \bm q) = \int\limits_{-\infty}^\infty \frac{d Q_z}{2 \pi} D(i \omega, \bm Q) \nonumber \\
	&& = -\frac{4\pi^2 Z^2 e^4}{\varepsilon_\infty^2 M \Omega_0}\frac{R_s}{\sqrt{(qR_s)^2+1}}\frac{1}{\omega^2+\omega_0^2} \, . \label{D2DSM}
\end{eqnarray}
This corresponds to the interaction in Eq.~(10) in the main text with the dimensionless EPI constant $\lambda_{ep}$,
\begin{eqnarray}
	&& \lambda_{ep} = -N_F U(0, 0) = \frac{2 \pi Z^2 e^4 m R_s}{\varepsilon_\infty^2 M \Omega_0 \omega_0^2} \, , \label{lambdaep}
\end{eqnarray}
where $N_F = m/(2 \pi)$ is the 2D density of states per spin, $m$ the effective mass.
The important result here is that $\lambda_{ep} \propto R_s$ which is used to plot Fig.~6 in the main text.

\section{$\Lambda$ and $\gamma_f$ for the EPI}

Here, we calculate $\Lambda$ and $\gamma_f$ [see Eqs.~(\ref{gf}), (\ref{LambdaSM})] for the EPI in Eq.~(\ref{D2DSM}).
First, we extract the finite-range (forward-scattering) component of the interaction,
\begin{eqnarray}
	&& \hspace{-10pt}  N_F U_f(i\omega, q) = N_F U(i \omega, q) - N_F U(i\omega, 2 k_F) \nonumber \\
	&&  \hspace{-10pt} = - \frac{\lambda_{ep}}{\sqrt{(q R_s)^2 + 1}} \frac{\omega_0^2}{\omega^2 + \omega_0^2} + \frac{\lambda_{ep}}{\sqrt{(2 z)^2 + 1}} \frac{\omega_0^2}{\omega^2 + \omega_0^2} \, , \label{Uf}
\end{eqnarray}
where $N_F = m/(2 \pi)$ is the 2D density of states per band, $z = k_F R_s$ is the dimensionless Fermi momentum, and $\lambda_{ep}$ is defined in Eq.~(\ref{lambdaep}).
The finite-range interaction is zero at $q = 2 k_F$ and has a maximum (in absolute value) at $q = 0$.
We only consider momentum transfers $q < 2 k_F$, as the scattering processes with $q > 2 k_F$ correspond to scattering off the Fermi surface which is irrelevant.
It is convenient to introduce the short-range coupling constant $\gamma_0$  (see Eq.~(12) in the main text),
\begin{eqnarray}
	&& \gamma_0 = - U(0, 2 k_F) N_F = \frac{\lambda_{ep}}{\sqrt{(2 z)^2 + 1}} \, .
\end{eqnarray}

Substituting Eq.~(\ref{Uf}) into Eq.~(\ref{gf}), we find the finite-range interaction coupling constant, 
\begin{eqnarray}
	&& \hspace{-13pt}\gamma_f = - \frac{2}{\pi v_F} \int\limits_0^\infty U_f(i \omega = 0, r) \, dr \nonumber \\
	&& = - \frac{2}{\pi v_F} \int\limits_0^\infty dr \, \int\limits_0^{2 k_F} \frac{q dq}{2 \pi} J_0(q r) U_f(0, q) \nonumber \\
	&& = - \frac{2}{\pi} \int\limits_0^2 N_F U_f(0, k_F u) \, du \, ,
\end{eqnarray}
where we integrate over momentum transfers $q \in (0, 2 k_F)$, $J_0(x)$ is the Bessel function and we changed the variable $q = 2 k_F u$ in the last line.
The integral over $u$ is elementary and results in Eq.~(11) in the main text.

At small temperatures $T \ll \omega_0$ the integral over $\tau$ in Eq.~(\ref{LambdaSM}) can be extended to the interval $(-\infty, \infty)$, i.e. we consider $\Lambda$ in the limit of $T = 0$.
In this case, it is also convenient to work with the zero-temperature Fourier transform $U_f(\tau, q)$,
\begin{eqnarray}
	&& N_F U_f(\tau, q) = \frac{\lambda_{ep} \omega_0 e^{-\omega_0 |\tau|}}{2\sqrt{(2 z)^2 + 1}} - \frac{\lambda_{ep} \omega_0 e^{- \omega_0 |\tau|}}{2 \sqrt{(q R_s)^2 + 1}} \, . \label{Utau}
\end{eqnarray}
$U_f(\tau, r)$ in Eq.~(\ref{LambdaSM}) is approximated by the Fourier transform for $q < 2 k_F$,
\begin{eqnarray}
	&& U_f(\tau, r) = \int\limits_0^{2 k_F} \frac{q dq}{2 \pi} J_0(qr) U_f(\tau, q) \, . \label{Utaur}
\end{eqnarray}
This results in the following expression for $\Lambda$,
\begin{eqnarray}
	&& \hspace{-30pt} \ln \Lambda + C_{\mathrm{E}} = \int\limits_{0}^\infty du \, J_0(u) \int\limits_0^{2 k_F} dq \,  \nonumber \\
	&& \hspace{15pt} \times  \int\limits_{-\infty}^\infty d\tau \, \frac{N_F U_f(\tau, q)}{\pi k_F \gamma_f} \ln\left(\frac{u^2}{(qv_F)^2} + \tau^2\right)\, , \label{Lu}
\end{eqnarray}
where $u = q r$.
As $U_f(\tau, q) \propto e^{-\omega_0 |\tau|}$ [see Eq.~(\ref{Utau})] we integrate over $\tau$ first using the following identity,
\begin{eqnarray}
	&& \frac{\omega_0}{2} \int\limits_{-\infty}^\infty d\tau \, e^{-\omega_0 |\tau|}  \ln \left( \frac{u^2}{(qv_F)^2} + \tau^2 \right) \nonumber \\
	&& = 2 \ln\left(\frac{u}{q v_F}\right)  + 2 \int\limits_0^\infty \frac{\cos t}{t + a} \, dt \, ,
\end{eqnarray}
where $a = u \omega_0/(q v_F)$.
This representation allows us to take the integral over $u$ using the following identities,
\begin{eqnarray}
	&& 	\int\limits_0^\infty J_0(u) \ln u \, du = - C_{\mathrm{E}} - \ln 2 \, ,\\
	&& \int\limits_0^\infty \frac{J_0 (u) \, du}{u + b} = \frac{\pi}{2} \left(H_0(b) - Y_0(b)\right) \, , \\
	&& \hspace{-10pt} \frac{\pi}{2} \int\limits_0^\infty \left(H_0(b t) - Y_0(b t)\right) \cos t \, dt = \frac{\ln\left(b + \sqrt{b^2 - 1}\right)}{\sqrt{b^2-1}} \, . \label{HY}
\end{eqnarray}
Here, $H_0(b)$ is the Struve function and $Y_0(b)$ is the Bessel function of second kind.
One way to prove Eq.~(\ref{HY}) is to use integral representations for $H_0(x)$ and $Y_0(x)$.
With this, Eq.~(\ref{Lu}) can be simplified to the following expression,
\begin{eqnarray}
	&& \hspace{-20pt} \ln \Lambda + C_{\mathrm{E}} = \frac{2 \lambda_{ep}}{\pi \gamma_f z} \int\limits_0^{b_0} db \, \left( \frac{1}{\sqrt{b^2 + b_1^2}} - \frac{1}{\sqrt{b_0^2 + b_1^2}}\right)\nonumber \\
	&& \times \left[\ln(2 \omega_0 b) + C_{\mathrm{E}} - b \frac{\ln\left(b + \sqrt{b^2 - 1}\right)}{\sqrt{b^2-1}} \right] \, , \label{B1int}
\end{eqnarray}
where $b = q v_F/\omega_0$, $z = k_F R_s$ is the dimensionless Fermi momentum, $b_0 = 4 E_F/\omega_0$, and $b_1 = v_F/(\omega_0 R_s)$.
Finally, we use the following identities,
\begin{equation}
	\int\limits_0^{b_0} \frac{d b}{\sqrt{b^2 + b_1^2}} = \mathrm{arcsinh}\left(\frac{b_0}{b_1}\right) = \mathrm{arcsinh}(2z)\, ,
\end{equation}
\begin{equation}
	\int\limits_0^{b_0}  \frac{\mathrm{arccosh}(b) b \, db}{\sqrt{b^2 - 1}} = \mathrm{arccosh}(b_0) \sqrt{b_0^2-1} -b_0 + \frac{\pi}{2} \, , 
\end{equation}
\begin{eqnarray}
	&& \hspace{-15pt} \int\limits_0^{2z} \frac{\ln v \, dv}{\sqrt{v^2 + 1}} = \frac{1}{2} \mathrm{arcsinh}^2(2 z) - \ln (2) \mathrm{arcsinh}(2 z) \nonumber \\
	&& \hspace{30pt} - \frac{\pi^2}{12} + \frac{1}{2} \mathrm{Li}_2\left(\left(\sqrt{4 z^2 + 1} - 2z\right)^2\right) \, ,
\end{eqnarray}
where $\mathrm{Li}_2(x)$ is the dilogarithm function and $\mathrm{arccosh}(x) = \ln(x + \sqrt{x^2-1})$ is the inverse hyperbolic cosine. Using Eqs.~(11), (12) in the main text, we arrive at the following expression for the cut-off that we use for numerical calculations, 
\begin{widetext}
	\begin{eqnarray}
		&&  \ln\left(\frac{\Lambda R_s}{v_F}\right) = \frac{4 \gamma_0}{\pi \gamma_f} \left(\frac{\sqrt{b_0^2 - 1}}{b_0} \ln \left(b_0 + \sqrt{b_0^2 - 1}\right) + \frac{\pi}{2b_0} - \ln(4 z) \right) \nonumber \\
		&& + \frac{2 \lambda_{ep}}{\pi \gamma_f z} \left(\frac{1}{2} \mathrm{arcsinh}^2(2 z) - \frac{\pi^2}{12} + \frac{1}{2} \mathrm{Li}_2\left(\left(\sqrt{4 z^2 + 1} - 2z\right)^2\right) - I_{\frac{1}{2}}\left(b_0, b_1\right)\right) \, , \label{Lambdanum}
	\end{eqnarray}
\end{widetext}
where we introduced the following notation,
\begin{eqnarray}
	&& I_{s}\left(b_0, b_1\right) = \int\limits_0^{b_0} d b \, \frac{b \ln\left(b + \sqrt{b^2 - 1}\,\right) \,} {\left(b^2 + b_1^2\right)^{s} \sqrt{b^2-1}} \, ,
\end{eqnarray}
for real $s > 0$.
We point out that the integrand is real-valued and independent of the branch of $\sqrt{b^2-1}$ at $b < 1$.

The isotope exponent $\alpha$ follows from Eq.~(\ref{Lambdanum}),
\begin{eqnarray}
	&& \alpha = \frac{1}{2} \frac{d \ln \Lambda}{d \ln \omega_0} = - \frac{\lambda_{ep} b_1^2}{\pi z \gamma_f} I_{\frac{3}{2}}(b_0, b_1) \nonumber\\
	&& + \frac{2 \gamma_0}{\pi \gamma_f} \left[\frac{\sqrt{b_0^2 - 1}}{b_0} \mathrm{arccosh}(b_0) + \frac{\pi}{2 b_0} - 1\right] \, . \label{alphaSM}
\end{eqnarray}

We can expand all terms in the limit $z \gg 1$. Approximate coupling constants simplify to the following expressions,
\begin{eqnarray}
	&& \gamma_0(z) \approx \frac{\lambda_{ep}}{2 z}\, , \hspace{5pt} \gamma_f (z) \approx \frac{2\lambda_{ep}}{\pi z} \ln\left(\frac{4z}{e}\right) \, . \label{gammaapp}
\end{eqnarray}
The large $z$ asymptotics of the integral $I_{\frac{1}{2}}(b_0, b_1)$, $b_0 = 4 E_F/\omega_0$, $b_1 = v_F/(\omega_0 R_s)$, strongly depends on $b_1$,
\begin{eqnarray}
	&& \hspace{-15pt} I_{\frac{1}{2}}(b_0, b_1) \nonumber \\
	&& \hspace{-10pt} \approx \left\{
	\begin{array}{cc}
		\frac{\displaystyle 1}{\displaystyle 2} \ln\left(4 z\right) \ln\left(\frac{\displaystyle 4 z^3}{\displaystyle \tilde{\omega}_0^2}\right) - \frac{\displaystyle \pi^2}{\displaystyle 12} + \frac{\displaystyle \pi \tilde{\omega}_0}{\displaystyle 2 z} \, , &  b_1 \gg 1 \, , \\
		\frac{\displaystyle 1}{\displaystyle 2} \ln^2 \left(b_0 + \sqrt{b_0^2 - 1}\right) + \frac{\displaystyle \pi^2}{\displaystyle 8} - \frac{\displaystyle \pi z}{\displaystyle 2 \tilde{\omega}_0} \, , & b_1 \ll 1\, .
	\end{array}
	\right.
\end{eqnarray}
Here, $\tilde{\omega}_0 = m R_s^2 \omega_0$ is the dimensionless phonon frequency.
This results in the following asymptotic behavior of the cut-off $\Lambda$ at $z \gg 1$,
\begin{equation}
	\frac{\Lambda}{\omega_0} \approx \left\{
	\begin{array}{cc}
		\exp\left[-\frac{\displaystyle \pi \tilde{\omega}_0}{\displaystyle 2 z\ln\left(4 z/e\right)}\right] \, , & v_F \gg \omega_0 R_s \\
		\frac{\displaystyle 2 z^{\frac{3}{2}}}{\displaystyle\sqrt{e} \tilde{\omega}_0} \exp\left[-\frac{\displaystyle f(b_0)}{ \displaystyle 2 \ln\left(4 z/e\right)}\right] \, , & v_F \ll \omega_0 R_s \, ,
	\end{array}
	\right.
\end{equation} 
where $f(b_0)$ is the following function,
\begin{eqnarray}
	&& \hspace{-10pt} f(b_0) = \left(\mathrm{arccosh}(b_0) - \frac{\sqrt{b_0^2 - 1}}{b_0}\right)^2 \nonumber \\
	&& \hspace{15pt}  + \frac{1}{b_0^2} - \frac{\pi}{b_0} + \frac{5 \pi^2}{12} - \frac{\pi z}{\tilde{\omega}_0} \, .
\end{eqnarray}
First, we note that $\Lambda \sim \omega_0$ at $v_F \gg \omega_0 R_s$, which is quite intuitive because $\omega_0$ plays the role of the smallest cut-off energy scale in the system.
In the opposite limit $v_F \ll \omega_0 R_s$, the cut-off $\Lambda < \omega_0$ crosses over to the power-law $\Lambda \propto z^{3/2}/(m R_s^2) \ll \omega_0$ at $1 \ll z \ll \sqrt{\tilde{\omega}_0}$.
This regime is accessible if $\omega_0 \gg 1/(m R_s^2)$.

We stress that if $\omega_0 \lesssim 1/(m R_s^2)$, then the regime where $b_1 \ll 1$ and $z \gg 1$ is no longer possible as these two conditions are mutually exclusive: $\tilde{\omega}_0 \gg z \gg 1$. Therefore, $b_1 \gtrsim 1$ and $\Lambda \sim \omega_0$. Hence, in this regime we expect the standard isotope effect with exponent $\alpha \approx 0.5$. Strong dependence of the cut-off on the electron density corresponds to the regime when $z \lesssim 1$ which is opposite to the semiclassical limit $z \gg 1$ considered in this paper.

Therefore, our most relevant results correspond to the parameter regime when $\omega_0 \gg 1/(m R_s^2)$. In this case, the regime $\tilde{\omega}_0 \gg z \gg 1$ is possible. In this regime, the cut-off $\Lambda$ becomes a function of the electron density $n_e$. In particular, $\Lambda \propto k_F^{3/2} \propto n_e^{3/4} \ll \omega_0$ at $\sqrt{\tilde{\omega}_0} \gg z \gg 1$. In this regime, $\Lambda$ is nearly independent of $\omega_0$ which results in a strongly suppressed isotope effect.

\section{Estimates for $R_s$}

Let us first estimate the Thomas-Fermi (TF) screening length $R_{\mathrm{TF}}$. For this, we approximate the polarization operator by the particle-hole bubble (Lindhard function),
\begin{eqnarray}
	&& \Pi(i \omega, q) = 2 \int \frac{d^3 k}{(2 \pi)^3} \frac{n_{\bm k + \bm q} - n_{\bm k}}{\xi_{\bm k + \bm q} - \xi_{\bm k} - i \omega} \, ,
\end{eqnarray}
where the additional factor of 2 is due to the spin degeneracy, $n_{\bm k}$ is the occupation number of the electron state with momentum $\bm k$, $\xi_{\bm k} = k_\parallel^2/(2m) - E_F$ is the electron dispersion and $k_\parallel^2 = k_x^2 + k_y^2$.
Here, we take into account that the Fermi surface is a cylinder as we consider a layered material with weak inter-layer hopping.
In the static limit $i \omega = 0$, $q \to 0$, we find,
\begin{eqnarray}
	&& \Pi_0 \equiv \Pi(0, q\to 0) = -2  \int \frac{d^3 k}{(2 \pi)^3} \delta\left(\xi_{\bm k}\right) \, ,
\end{eqnarray}
where $\delta(x)$ is the delta function.
As $\xi_{\bm k}$ does not depend on $k_z$, we integrate over the entire interval $k_z \in [-\pi/a_\perp, \pi/a_\perp]$, where $a_\perp$ is the inter-plane distance. The remaining integral over $k_x$ and $k_y$ is elementary,
\begin{eqnarray}
	&& \Pi_0 = - \frac{m}{\pi a_\perp} \, .
\end{eqnarray}
The Coulomb kernel in Eq.~(\ref{eph}) can now be dressed by the static polarization bubble as follows,
\begin{eqnarray}
	&& V_C (\bm Q) = -\frac{4 \pi Ze^2}{\varepsilon_\infty Q^2} \frac{1}{1 - \Pi_0 V_0(\bm Q)} \, , \label{TF}\\
	&&  V_0(\bm Q) = \frac{4 \pi e^2}{\varepsilon_\infty Q^2} \, ,
\end{eqnarray}
where $V_0(\bm Q)$ is the electron-electron Coulomb interaction.
Comparing Eq.~(\ref{TF}) with Eq.~(\ref{VCQ}), we find $R_s = R_{\mathrm{TF}}$, with
\begin{eqnarray}
	&& R_{\mathrm{TF}} = \frac{1}{2} \sqrt{a_\perp a_B} \, ,
\end{eqnarray}
where $a_B = \varepsilon_\infty/(m e^2)$ is the effective Bohr radius. 
The TF approximation is valid if $R_{\mathrm{TF}} \gg \lambda_F$ which is not satisfied even in 3D materials with large $k_F \sim 1\,$nm$^{-1}$.

A better estimate follows from the relation between the speed of longitudinal acoustic phonons, $c_{\mathrm{ac}}$, and $R_s$, see Eq.~26.5 in Ref.~\cite{ashcroft1976},
\begin{eqnarray}
	&& R_s \approx \frac{c_{\mathrm{ac}}}{\omega_{ip}}, \hspace{5pt} \omega_{ip} = \sqrt{\frac{4 \pi Z e^2 n_e}{M \varepsilon_\infty}} \, ,
\end{eqnarray}
where $\omega_{ip}$ is the ionic plasma frequency, $n_e = Z/\Omega_0$ the electron density, $Z$ the charge per unit cell, and $M$ the mass of the unit cell.
Let us estimate $R_s$ in YBCO near the optimal doping where  $c_{\mathrm{ac}} \approx 5 \times 10^3\,$m/s \cite{elbaumAcousticStudiesHigh1996}.
In order to estimate $\omega_{ip}$, we require the mass of the YBCO unit cell, $M \approx 1.2 \times 10^6 \, m_0$, the dielectric constant, $\varepsilon_\infty \approx 3.8$ \cite{loboUnexpectedBehaviourIR1995,castillo-lopezNearfieldRadiativeHeat2020}, the unit cell volume $\Omega_0 = 173 \, \AA^3$, and the charge of the unit cell $Z \approx 0.66 \times 2$, where $0.66$ is the charge per Cu(2) atom \cite{mincheongFirstPrincipleCalculation2024}.
From this, we find $\omega_{ip} \approx 1.7 \,$meV and $R_s \approx 2\,$nm.
Optimal doping in YBCO corresponds to $k_F^* \approx 2.6 \,$nm$^{-1}$, hence $z^* = k_F^{*} R_s \approx 5$, which can still be considered as a large parameter, $z^* \gg 1$.

\end{document}